# A rational design of microwire/CNT hybrid fiber for tunable microwire metacomposites

D. Estevez, F.X. Qin[1]  Y. Luo, X. Zheng, H. Wang and H-X Peng

Institute for Composites Science Innovation (InCSI), School of Materials Science and Engineering, Zhejiang University
38 Zheda Road, Hangzhou, 310027, PR. China

[1] Corresponding author: faxiangqin@zju.edu.cn, Tel: + 86571 87953052.

# Abstract

Metacomposites are a type of engineering composite materials with metamaterial properties. We propose a design of hybrid filler consisting of ferromagnetic microwire and carbon nanotube to enable a tunable microwire metacomposite. There shows a sophisticated dependency of plasma frequency and bandwidth of double negative region on the carbon nanotube coating. Factors such as coating thickness, uniformity, morphology and number of microwires conspire to the formulation of the double negative characteristics. The intricate role of CNT on modulating negative $\varepsilon$ and $\mu$ is elucidated by the plasma frequency formula and split-close paths model based on the layer-like structure of CNT coating.

Metacomposites, defined as composites materials with metamaterials properties, is firstly proposed in the field of mechanical metamaterials [1-3] and recently draw much research interest in the field of electromagnetic metamaterials.[4,5] Explicitly explained by Luo and Qin et al. [5-7] metacomposites differ from the conventional metamaterials at three points: (i) metacomposites are 'true' composite materials rather than structures. (ii) metacomposites can be fabricated by conventional composites manufacture technology instead of micro-/nano- fabrication. (iii) the ultimate properties of metacomposites are not only dependent on the geometrical parameters but also dependent on the filler composition and properties. Following Carbonell and Liu et al.'s works [8-10] demonstrating the feasibility of utilizing ferromagnetic microwires for realizing metamaterials, our group has devoted enormous efforts to developing polymer metacomposites built upon such microwires and presents several types of configurations such as single-parallel, orthogonal arrangement and combination of different types of microwires.[6,7,11,12] Most recently, Chiriac and co-workers [13-15] also reported the double negative (DNG) behavior of several types of microwire media. With all these studies successfully demonstrating the metamaterial behavior, there lacks adequate control of the double negative features in terms of e.g., the negativity bandwidth. In this context, we consider here to further functionalize the microwire by introducing another phase, i.e., the carbon nanotube (CNT) to increase the freedom of tuning electromagnetic DNG features in microwire metacomposites.

Electrophoretic deposition (EPD) [17] was used to coat multiwall CNTs (Cheaptube, purity 95%) on the surface of melt-extracted amorphous $Co_{68.15}Fe_{4.35}Si_{12.25}B_{15.25}$ microwires with diameter of 70μm.[16] To prepare the CNT solution for the EPD process, CNTs were chemically oxidized in a mixture of $H_2SO_4$, $KMNO_4$, $NaNO_3$, and $H_2O_2$ to introduce carboxyl acid groups on the CNT surface and provide them with a negative surface charge (Fig.1a). The processed CNTs did not indicate any visual signs of agglomeration (Fig.1b). The concentration of the CNTs in the final solution was slightly less than 1 mg/ml. The process illustrated in Fig.1c was carried out at 10V for 60 and 120 seconds, and at 20V for 60 seconds to yield different coating thicknesses (1.73 ±0.30μm; 2.81±0.11μm for 10V series and 3.11 ±0.26μm for 20V). The morphology and topography of the coatings were examined by scanning electron microscopy (SEM) in a Hitachi S-4800 cold field emission scanning electron microscope and a Bruker icon atomic force microscope (AFM), respectively. The composites were prepared by aligning a different number (1, 3, 7) of CNT/microwire hybrid fibers in a parallel manner between two facing molds with a fixed fiber spacing $s$ of 3mm. Samples with dimensions of 22.86mm × 10.16mm × 5mm were realised after cured (Fig.1d) at 125°C for 20mins. Scattering parameters of the composites were measured with Rohde & Schwarz ZNB 40 vector network analyzer by using a WR-90 waveguide in $TE_{10}$ dominant mode from 8.2 to 12.4GHz. Complex permeability μ and complex permittivity ε were then determined from the measured $S_{11}$ and $S_{21}$ parameters using the Nicolson–Ross–Weir algorithm[18].

In samples treated with 10V-120s and 20V-60s the coating appeared to be more homogeneously dispersed over the whole surface of the wire when compared to the sample treated at 10V-60s (Fig.1e). A close-up view of the surface of the fibers shows disordered distribution of the CNT with numerous inter-tangled bundles. The CNTs deposited on the wire surface contribute to the increased roughness, i.e., many obvious convex hills were present on the hybrid fiber surfaces (Fig.1f), which can enhance interfacial strength by mechanical interlocking between the fibers and polymer matrix.

Figure 2a shows the frequency dependence of the measured transmission parameters $S_{21}$ for sample containing 1, 3 and 7 hybrid fibers, respectively, which show no readily identifiable features associated with negative index. We then turn to the inversion of the scattering equations leading to the following form that connects refractive index to $S_{21}$[19]:

$$\cos(nkd) = \text{Re}\left(\frac{1}{S_{21}}\right) - \frac{1}{2|S_{21}|^2}\left(A_1 S_{11} + A_2 S_{21}\right) \quad (1)$$

where $k$, $n$, and $d$ are the wavenumber, refractive index and maximum length of the unit element, respectively; $A_1$ and $A_2$ are real-valued functions that tend to be zero in the absence of losses. Eq. (1) points to a strong correlation between the phase of $S_{21}$ and $n$. The dip in the phase or phase reversal of $S_{21}$ is an indicator of the existence of double negative region.[19] Therefore, the shaded frequency regions in Fig.2b mark double negativity for the samples containing 3 and 7 hybrid fibers. However, such phase reversal is absent for composites containing a single hybrid fiber regardless of the CNT thickness coating, suggesting that a left-handed property cannot be realized in extreme dilute composites due to extreme small electromagnetic excitation.

The direct observation of double negative region is from the permittivity and permeability spectra as shown in Figure 3a-c. In Fig. 3a, for composites containing more than one uncoated wire, a "natural" metacomposite feature, i.e., without the need of any biasing fields can be appreciated over the frequency region of 10-11GHz. However, such a DNG characteristic disappears in the metacomposite containing a single wire. This observation is in agreement with our previous analysis.[7] Notably, in Fig.3b and c, one should notice that the double negative region identified from the ε and μ spectra is consistent with that obtained from the phase spectra. To further elucidate the CNT effect, we plot in Fig.4 the variations of the double negative region, plasma frequency and bandwidth as the function of CNT coating thickness. There shows complex dependencies of plasma frequency and double negative bandwidth on the CNT coating thickness and they are not mutually consistent. Note that coating the wire with a CNT thickness of 1.73μm proves to be effective in widening the DNG region of the amorphous wire for compositing containing 3 fibers. By contrast, in the case of 7 fibers, the plasma frequency and DNG bandwidth of the uncoated wire are less affected by the thickness of CNT coating.

Now let us address the exact roles of the CNT on the metamaterial properties of our samples. The roles of CNT lie threefold: thickness, uniformity, and morphology. We begin with the discussion on the case of three fibers when the long-range wire interaction can be neglected. First, the thickness will be the only concern if the coating is perfectly uniform and smooth. In that case, the plasma frequency will be increased due to the increase of wire diameter according to $f_p^2 = \frac{c^2}{2\pi b^2 \ln\left(\frac{s}{a}\right)}$, which is determined by the wire radius *a* and spacing *s*[20]. This can explain the increased plasma frequency for composite containing fibers with 1.73μm coating as compared to the uncoated wire. Nevertheless, it is not applicable as the coating thickness increases further; this is where we need to consider the uniformity and morphology. As indicated in the cross-sectional view of the hybrid fibers (Fig.5a), with increasing thickness, the coating tends to become layer-wise and hence the interfacial polarization between layers is enhanced. As a result, the plasma frequency is downshifted. The inconsistency between the changing trend of plasma frequency and DNG bandwidth with the CNT coating thickness from 2.81 to 3.11μm should be attributed to the magnetic permeability. Compared to negative permittivity, negative permeability are more difficult to build up. Although the magnetic responses mainly come from the magnetic microwires in the present case, the CNT coating has a significant influence on the permeability. 1.73μm coating shielded the otherwise strong magnetic response due to the functionalized CNT non-uniformly coated on the wire.[21] With increasing coating thickness, the permeability is significantly enhanced, which can be explained in the framework of 'split closed path' theory [22]. With reference to Fig.5a, in the *x-y* plane, the CNT coating consists of numerous inter-tangled bundles with a disordered arrangement. These CNT bundles construct closed paths. Although the path is closed in the *x-y* plane, splits exist between bundles in *x-z* plane. On the other hand, in the waveguide measurement, the magnetic field *H* has a component perpendicular to the plane of these split closed paths, while electrical field *E* is parallel to split closed paths, as shown in Fig.5b. The quantity of split closed paths through the thickness direction is restricted for a single layer. Multilayer CNT coating is "easier to behave" negative μ' than monolayer CNT coating due to strong coupling between layers 21 which is proved by our results when the thickness of the CNT coating is 3.11μm. In spite of similar thickness, the number of closed-paths is reduced in 2.81μm thickness sample (Fig.5a), which accounts for its less negative permeability values in comparison with 3.11μm thickness sample. In summary, the influences of CNTs coating on magnetic responses of composites are mainly twofold: (i) a great number of split closed paths consisting of CNT bundles enhance electrical excitation; (ii) incident magnetic field induces a magnetic component perpendicular to these closed paths. Finally, the electromagnetic behavior of the sample containing 7 fibers is much less influenced by CNT coating due to the strong wire-wire interaction induced in this configuration among the closely packed microwires[23].

To recapitulate, we have designed and fabricated polymer-based composites containing CNT/microwire hybrid fibers in a parallel fashion. Varying the thickness and structural features of the CNT coating prove to be efficient in tuning the metamaterial property of the composites containing hybrid fibers. The layer-like structure of CNT coating in tandem with the plasma frequency formula and split-close paths model explain the role of CNT in modulating negative ε and μ. From such analysis, it is established that a multilayer coating structure is in favor of overall double negative property. Our findings afford a convenient and feasible strategy to realize tunable metacomposites, which can meet a spectrum of microwave device applications such as absorbers, phase shifters, microwave cloaks and sensors.

**Acknowledgements**

This work was supported by NSFC No. 51501162 and No. 51671171, Aeronautical Science Foundation GFJG-112207-E11502 and National Youth Thousand Talent Program' of China.The authors wish to thank Mr. Jiabin Xin for his valuable input on the data analyses.

**Figure Captions:**

Fig.1 (color online): Preparation of the CNT/wire hybrid fibres and their composites: (a) Chemical functionalization of CNTs; (b) dispersed functionalized CNT solution; (c) EPD deposition on wire surface; (d) configuration of composites containing CNT/wire hybrid fibres. Effect of different deposition parameters on the morphology and topology of the CNT coating: (e) side-view SEM images for the CNT coating; (f) AFM images of the same set of samples.

Fig.2 (color online) (a) Effect of CNT coating on the magnitude of scattering parameter $S_{21}$ for composites containing 1, 3 and 7 hybrid fibers in a parallel manner; (b) effect of CNT coating on the phase of $S_{21}$, for composites containing 1, 3 and 7 hybrid fibers in a parallel manner. The shaded regions correspond to the left-handed propagation regime.

Fig.3 (color online) (a) $\varepsilon'$, $\mu'$ spectra for composites containing different numbers of uncoated wire, the inset shows the enlarged regions where $\varepsilon'$, $\mu'$ simultaneously exhibit negative values; (b) $\varepsilon'$ spectra for composites containing different number of hybrid fibers; (c) $\mu'$ spectra for composites containing different number of hybrid fibers.

Fig.4 (color online) (a) Dependencies of CNT coating thickness on the double negative (DNG) frequency region, (b) DNG bandwidth and (c) plasma frequency for composites containing 3 fibers and 7 fibers.

Fig. 5 (color online) Illustrations for (a) split-close path model corresponding to the SEM images of hybrid fibers with different CNT coating thickness and (b) propagation of $TE_{10}$ wave in the rectangular waveguide.

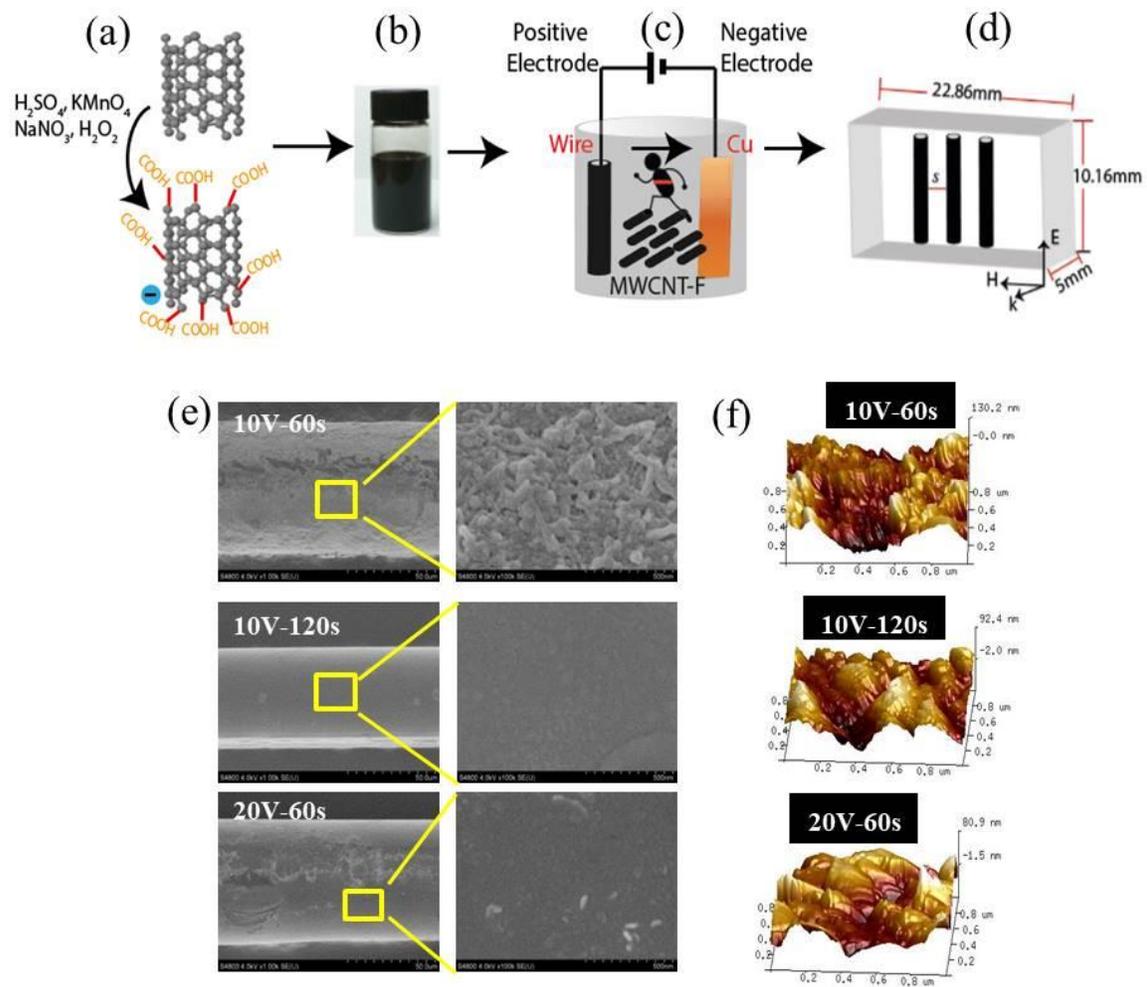

**FIG 1 Qin and Estevez**

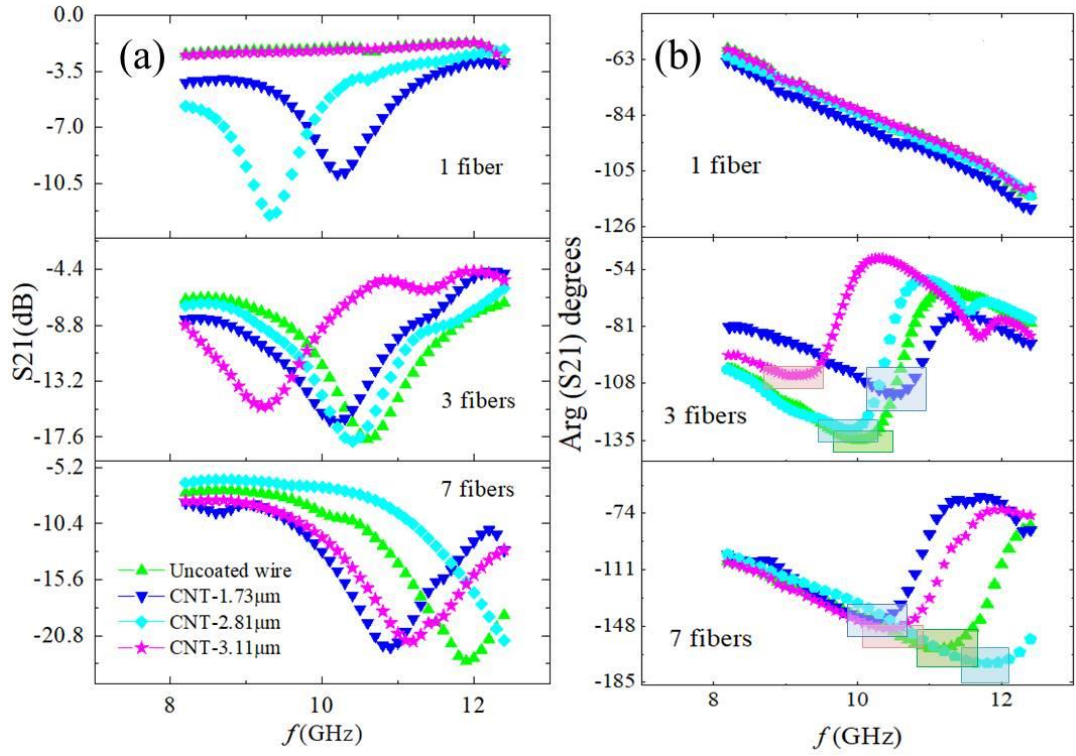

**FIG 2 Qin and Estevez**

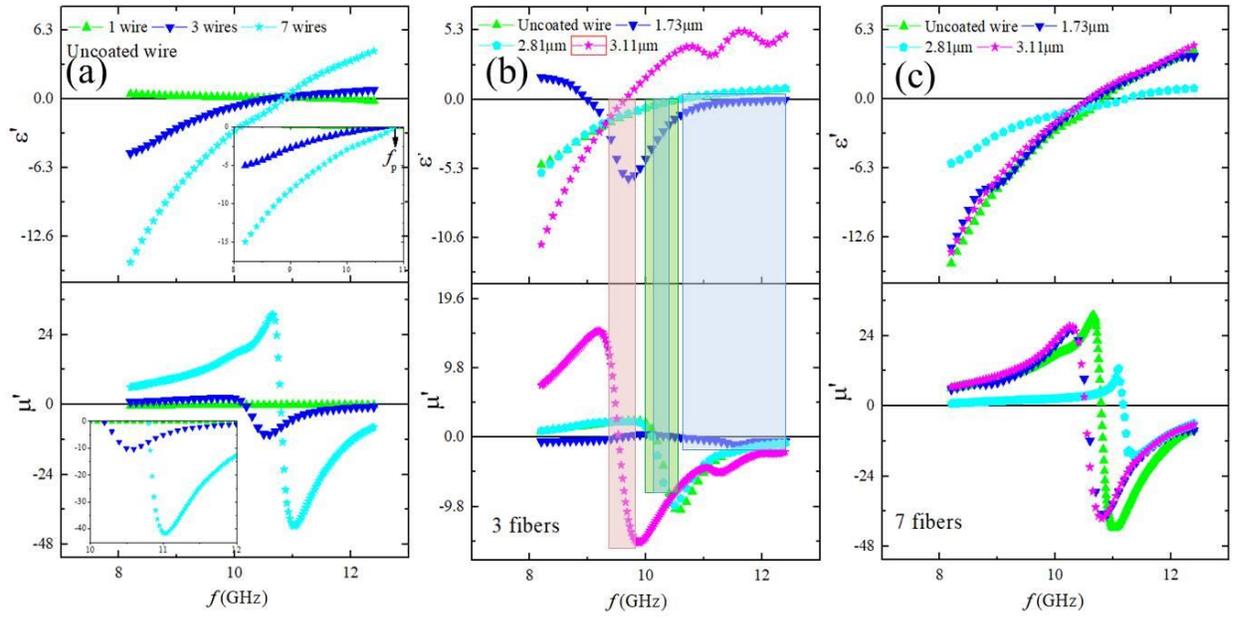

**FIG 3 Qin and Estevez**

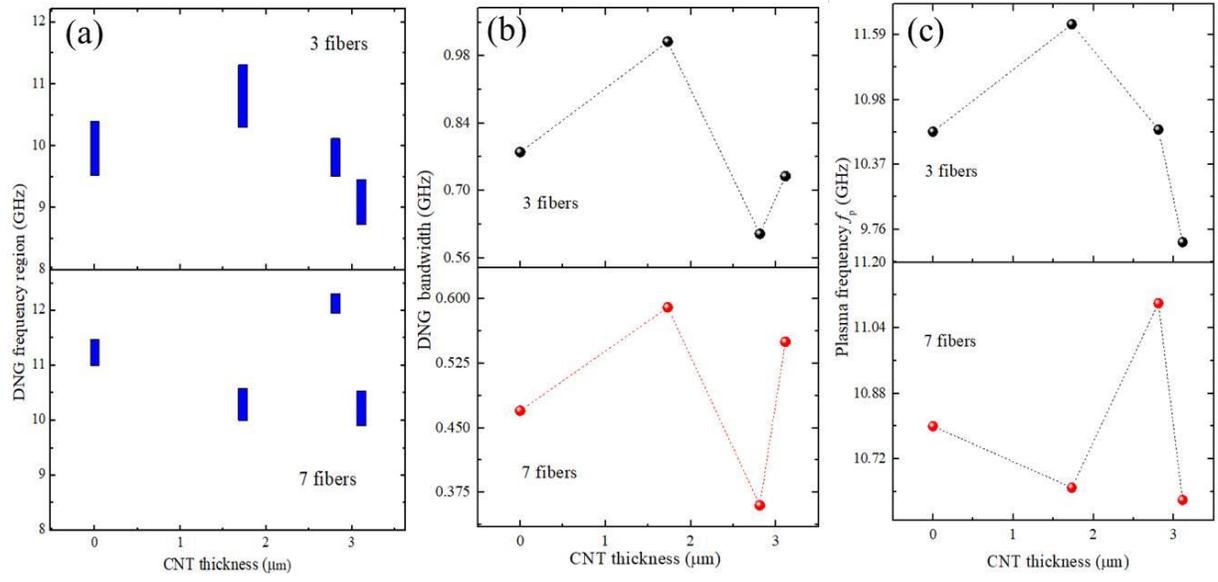

**FIG 4 Qin and Estevez**

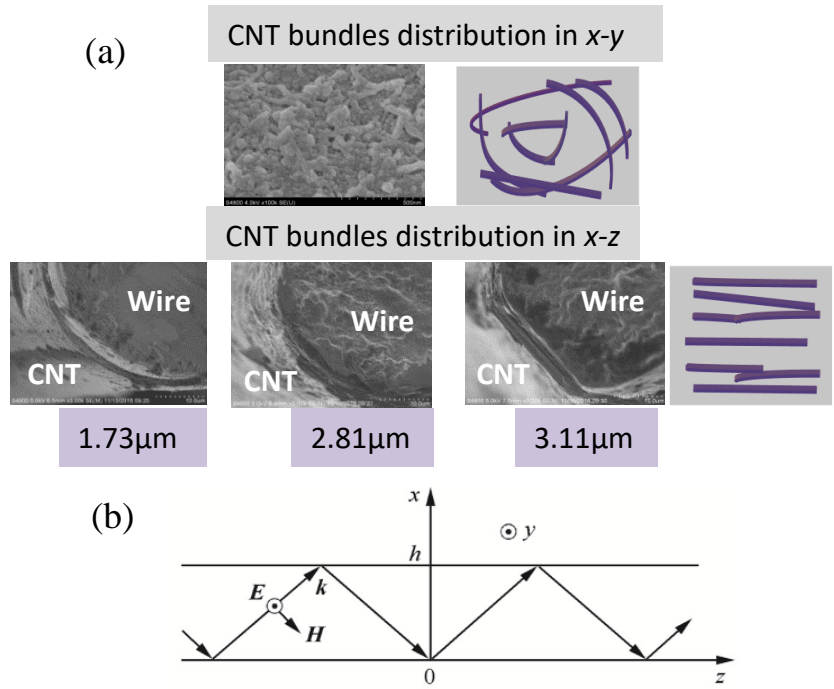

**FIG 5 Qin and Estevez**